# A Rewriting Logic Approach for Automatic Composition of Web Services

Walid Berrouk[1] and Ouanes Aissaoui[2]

[1] LIRE Laboratory, Mentouri University, P.O. Box 325, City Ain El Bey 25017 Constantine, Algeria

[2] LISCO Laboratory, Badji Mokhtar-Annaba University, P.O. Box 12, 23000 Annaba, Algeria

**Abstract**

Nowadays, Web Services (WS) remain a main actor in the implementation of distributed applications. They represent a new promising paradigm for the development, deployment and integration of Internet applications. These services are in most cases unable to provide the required functionality; they must be composed to provide appropriate services, richer and more interesting for other applications as well as for human users. The composition of Web services is considered as a strong point, which allows answering complex queries by combining the functionality of multiple services within a same composition. In this work we showed how the formalism of graphs can be used to improve the composition of web services and make it automatic. We have proposed the rewriting logic and its language Maude as a support for a graph-based approach to automatic composition of web services. The proposed model has made possible the exploration of different composition schemas as well as the formal analysis of service compositions. The paper introduces a case study showing how to apply our formalization.

## 1. Introduction

The service-oriented architecture [1] is a new paradigm that aims to build software systems using basic loosely coupled services. These services are in most cases unable to provide the required functionalities; they must be composed to provide appropriate services, richer and more interesting for other applications as well as for human users.

Automatic composition of web services has drawn a great deal of attention recently. By composition, we mean taking advantage of currently existing web services to provide a new service that does not exist on its own [2]. Therefore, in order to have a more complex service we can use some semantically related simpler web services and execute them in such a way that the whole set provides the desired service. Service composition is usually defined using two complementary approaches: the choreography and orchestration. In orchestration [3, 16, 17], the involved web services are under control of a single endpoint central process (another web service). This process coordinates the execution of different operations on the Web services participating in the process. The invoked Web services neither know and nor need to know that they are involved in a composition process and that they are playing a role in a business process definition. Only the central process (coordinator of the orchestration) is conscious of this aim, thus, the orchestration is centralized through explicit definitions of operations and the invocation order of Web services. Choreography [3, 16, 17], in contrast, does not depend on a central orchestrator. Each Web service that participates in the choreography has to know exactly when to become active and with whom to interoperate. Choreography is based on collaboration and is mainly used to exchange messages in public business processes. All Web services which take part in the choreography must be conscious of the business process, operations to execute, and messages to exchange as well as the timing of message exchanges.

Service composition has been addressed by several researches. The study of existing literature shows that the problem of automatic composition of web services is inherently very difficult because the data are unstable and the Web is dynamic.

Different formalisms have been proposed for the web services composition by several research teams around the world. Among these formalisms we can mention: graphs [2], Petri nets [4, 15, 18], process algebras [5], finite state machines [14] and UML [13]. To the difference of these approaches, our contribution in this work is to show how the formalism of graphs can be used to improve the composition of web services and make it automatic. More specifically we aim to formalize by using rewriting logic and its Maude system the graph-based algorithm of web services composition presented in [6].

The remainder of the paper is organized as follows: In Section 2, the rewriting logic and Maude language are briefly introduced. Section 3 details the proposed approach. Section 4 provides an example of how our approach works.

Finally, section 5 contains a brief conclusion and describes the future plans.

## 2. Basic concepts

2.1 Rewriting logic

The rewriting logic is a logic of concurrent changes which can treat the state and the computing of the concurrent systems. It was introduced by Meseguer [8] as a consequence of work on the general logics. Consequently, this logic was largely used to specify and analyse systems and languages in various applicability. Thus the logic of rewriting offers a formal framework necessary for the specification and the study of the behaviour of the concurrent systems. Indeed, it makes it possible to reason on possible complex changes corresponding to the atomic actions axiomatized by the rewriting rules. The key point of this logic is that the logical deduction, which is intrinsically concurrent, corresponds to computing in a concurrent system [9, 10].

In this logic the static aspect of the systems is represented by a subjacent logic called membership equational logics. The dynamic aspect is represented by rewriting theories describing the possible transitions between the states of the concurrent system [7]. The equational logic makes it possible for us to carry out modular specifications.

The rewriting logic is proposed as a logical framework in which other logics can be represented, and as a semantic framework to specify several systems and languages in varied fields. It offers techniques of formal analysis making it possible to prove properties of the system to be specified, and to reason on its changes.

2.2 Maude

Maude [12] is a specification and programming language and also a high level system based on the rewriting logic. It implements and concretises the various concepts of the rewriting logic. Maude is simple, expressive and efficient. Maude offers few syntactic constructions and a well defined semantics.

It is, in addition, possible to describe naturally various types of applications. Maude is a language which supports easily the rapid prototyping and represents a programming language with competitive performances. In the Maude language, two levels of specification are defined. A first level relates to the specification of the system while second relates to the specification of the properties [11, 12]. The Maude programs are a rewriting theories and concurrent computing in Maude represent deductions in the rewriting logic.

This language was largely influenced by the language OBJ3, more precisely the equationnal part of Maude included OBJ3 as a sub-language.

## 3. Approach

Our objective in this work is to propose the rewriting logic [10] through the Maude language [12, 13] as a support for a graph-based approach to automatic composition of web services. The proposed implementation for this approach makes it possible to explore different schemas of composition as well formal analysis of service compositions using the tools built around the Maude such as its LTL model-checker. The purpose of this section is to present the formalization of the different phases of our algorithm for automatic compositions inspired by [6].

For better understanding this work, we start this section by introducing some basic concepts.

**Web Service**: a web service $S_i \in S$ (where S is the set of all services) is defined by a triplet (*ServiceName, InputTypeList, OutputTypeList*) where *ServiceName* indicates the name of the service, *InputTypeList* is the set of the elements in the input and *OutputTypeList* is the set of the results provided by this service (see Figure 1).

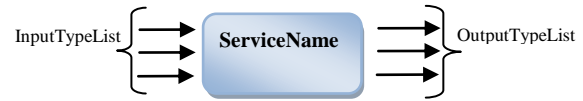

Fig. 1. Structure of an Atomic Web service.

**Composition of two services:** two services $S_i$ and $S_j$ can be composed if and only if the intersection of the outputs of $S_i$ with the inputs of $S_j$ is not empty.

Affinity: For any couple of services *A* and *B* ($A \neq B$), the association degree of *A* with *B* is defined as follows:

$$\text{aff}(A, B) = \frac{|A.\text{OutputTypeList} \cap B.\text{InputTypeList}|}{|B.\text{InputTypeList}|}$$
$$(0 \leq aff(A,B) \leq 1).$$

**Algorithm of composition model construction:**

**Inputs:**
```
The set of services S, S={S₁,…, Sₙ}
The initial node S_init = (T^i_Sinit ← S_init.InputTypeList, T^o_Sinit ←
   S_init.OutputTypeList)
```
**Outputs:** M (the composition model associated with S)

```
1:  Begin
2:    For each service B ϵ {S-S_init} do
3:      For each node A ϵ M do
4:        T^i_B ← ∅;   T^o_B ← ∅;
5:        if  T^o_A ∩ B.InputTypeList ≠ ∅ and B.OutputTypeList ⊄ T^o_A then
6:          M ← M ∪ (A, B);
7:          T^o_B ← T^o_A ∪ B.OutputTypeList;
8:          T^i_B ← T^i_A ∪ B.InputTypeList;
9:        end_if;
10:     end_For;
11:   end_For;
12: End.
```

This function is applied if and only if one service *A* can call a service *B*.

Our formalization of the graph-based approach to Web services composition inspired by [6] follows two principal steps. The first step consists to deriving the graph model using the module « COMP-MODEL » (implementation of the equational theory). This module offers the adequate semantic mechanisms specifying the constraints of connection between web services .The second step consists to generating the various planes of possible compositions making it possible to achieve a particular request. This phase is based on a mapping between the required elements and the provided elements by each node of the graph. The second part constitutes the dynamic aspect of the formalization. It will thus be implemented using a rewriting theory (the system module « COMPOSITION-PLAN »).

### 3.1 Construction of the graph-based compositions model

The graph-based services composition model presented in this section offers a description of the association between the web services components. In this model, the services filling the same functionalities cannot be included simultaneously because the required functionality can be accomplished before even as these services are not considered. Then an atomic service Si (Si ∈ S) can be includes in the model M if and only if Si achieves new functionalities.

The graph-based services composition model formalized in this work is made up of a set of nodes and edges between these nodes. Each node of the graph is equipped with two sets $T^i_A$ and $T^o_A$, these sets represent respectively the current elements of the input and the current provided results.

We propose the algorithm above for generating the graph-based model of a set of services starting from an initial node.

In this algorithm, an $T^o_A$ ∩ *B.InputTypeList* ≠ ∅ means that *B* can be called upon, *and B.OutputTypeList* ⊄ $T^o_A$ means that the invocation of *B* achieves new functionalities.

In order to formalize this algorithm, we propose the four Maude modules: «GRAPH», «SERVICE-SPEC», «SET-OPERATION» and «COMP-MODEL» respectively presented in the figures 2 to 5. For more clearness, we felt it important to give these theories by using the Maude code.

The Maude functional module « GRAPH » formalizes the graph data structure (see figure 2). In this module, after the declaration of the sorts and the relations of sub-sorts useful to describe the elements which a graph can contain, a set of operations and algebraic equations are introduced to specify the actions of addition of nodes and arcs to a graph.

To specify formally the concept of web service we suggest the functional module « SERVICE-SPEC » (figure 3). The most significant operation in this module is: *« op `(_:_->_`) : ServiceN TypeSet TypeSet -> Service [ctor prec 23] »* it is used to define the structure of atomic web services, the operation *« op __ : SetServiceN SetServiceN -*

> *SetServiceN [ctor id: none prec 25 ] . »* is used to generate a set of services definitions.

```
fmod GRAPH is
1)  sorts Node Nodes Edge Edges Graph .
2)  subsort Node < Nodes .
3)  subsort Edge < Edges .
4)  op niln : -> Node [ctor] .
5)  op __ : Nodes Nodes -> Nodes [ctor
    assoc comm id: niln prec 23] .
6)  op _in_ : Node Nodes -> Bool .
7)  op link`(_._`) : Node Node -> Edge
    [ctor prec 22] .
8)  op nile : -> Edge [ctor] .
9)  op __ : Edges Edges -> Edges [assoc
    comm id: nile prec 23] .
10) op _in_ : Edge Edges -> Bool .
11) op |_`,_| : Nodes Edges -> Graph
    [ctor prec 24 ] .
12) ops source target : Edge -> Node .
13) op addN : Graph Nodes -> Graph .
14) op addE : Graph Edges -> Graph .
15) vars e1 e2 : Edge .
16) vars n1 n2 : Node .
17) vars Ns Ns' : Nodes .
18) vars Es Es' : Edges .
19) eq n1 in n1 Ns = true .
20) eq n1 in Ns = false [owise] .
21) eq e1 in e1 Es = true .
22) eq e1 in Es = false [owise] .
23) eq source(link( n1 . n2 )) = n1 .
24) eq target(link( n1 . n2 )) = n2 .
25) eq addN(| Ns , Es | , niln) = | Ns , Es | .
26) eq addN(| Ns , Es | , n1 Ns' ) = if not (
    n1 in Ns ) then addN(| n1 Ns , Es | , Ns'
    ) else addN(| Ns , Es | , Ns' ) fi .eq
    addE(| Ns , Es | , nile ) = | Ns , Es | .
27) eq addE(| Ns , Es | , e1 Es') =  addE(
    addN(|Ns , e1 Es |, source(e1)
    target(e1)), Es') .
endfm
```

Fig. 2. The functional Module «GRAPH».

```
fmod SERVICE-SPEC is
1)  sorts Service ServiceSet Type
    TypeSet ServiceN SetServiceN .
2)  subsort Type < TypeSet .
3)  subsort Service < ServiceSet .
4)  subsort ServiceN < SetServiceN .
5)  op __ : TypeSet TypeSet -> TypeSet
    [ctor comm id: none prec 22 as soc] .
6)  op __ : SetServiceN SetServiceN ->
    SetServiceN [ctor id: none prec 25] .
7)  op `(_:_->_`): ServiceN TypeSet
    TypeSet -> Service [ctor prec 23] .
8)  op _;_ : ServiceSet ServiceSet ->
    ServiceSet [ctor comm id: none prec
    24 assoc] .
9)  op none : -> SetServiceN [ctor] .
10) op none : -> ServiceSet [ctor] .
11) op none : -> TypeSet [ctor] .
endfm
```

Fig. 3. The functional Module «SERVICE-SPEC».

```
fmod SET-OPERT is
1)  including SERVICE-SPEC .
2)  op Intersect : TypeSet TypeSet ->
    TypeSet [ctor] .
3)  op Union : TypeSet TypeSet -> TypeSet
    [ctor] .
4)  op Inclus : TypeSet TypeSet -> Bool
    [ctor] .
5)  op InputTypeList : Service -> TypeSet
    [ctor] .
6)  op OutputTypeList : Service ->
    TypeSet [ctor] .
7)  vars st1 st2 st3 st4 st5 st6 st7 st8
    st9 : TypeSet .
8)  vars t1 t2 t3 : Type .
9)  var ServiceN : ServiceN .
10) eq Intersect( t1 st1 , t1 st2) = t1
    Intersect ( st1 , st2 ) .
11) eq Intersect( st1 , st2 ) = none
    [owise] .
12) eq Union( t1 st1 , t1 st2 ) = t1
    Union ( st1 , st2 ) .
13) eq Union( t1 st1 , t2 st2 ) = t1 t2
    Union ( st1 , st2 ) .
14) eq Union( none , st2 ) = st2 .
15) eq Union( st1 , none ) = st1 .
16) eq Union( none , none ) = none .
17) eq Inclus( st1 , st1 st2 ) = true .
18) eq Inclus( st1 , st2 ) = false
    [owise] .
19) eq InputTypeList((ServiceN : st1 ->
    st2)) = st1 .
20) eq OutputTypeList(( ServiceN : st1 ->
    st2 )) = st2 .
21) eq Remove( t1 st1 ,  t1 st2 ) =
    Remove( st1 , st2 ) .
22) eq Remove( st2 , none ) = st2  .
23) eq Remove( st2 , st1 ) = st2 [owise].
endfm
```

Fig. 4. The functional Module «SET-OPERTION».

The functional module « SET-OPERTION » formalizes the different ensemblists operators and the two operations allowing to extract the elements in the input and the results provided starting from a services definition (Figure 4). In this module, we import firstly in mode "Including" the «SERVICE-SPEC» module already presented. Then, we give the signature of the ensemblists operators as well as their definitions using algebraic equations.

Lastly, we propose the last functional module «COMP-MODEL» formalizing the algorithm of composition model construction presented previously (see figure 5). This module, directly imports the two modules «GRAPH» and «SET-OPERT», and by transitivity the module «SERVICE-SPEC». In this module, the last conditional equation allows adding nodes to the graph after checking the two conditions $T_A^o \cap B.InputTypeList \neq \emptyset$ and $B.OutputTypeList \not\subset T_A^o$.

```
mod COMP-MODEL is
1)  including GRAPH .
2)  including SET-OPERT .
3)  sorts NodeSer Cmp-Mod N .
4)  subsort NodeSer < Node .
5)  op _&_ : ServiceSet Graph -> Cmp-Mod
    [ctor prec 25] .
6)  op <_`,_`,_> : ServiceN TypeSet
    TypeSet -> NodeSer [ctor prec 22].
8)  op NamSer : Node -> ServiceN [ctor].
9)  op NamSer : Service -> ServiceN [ctor].
10) vars ts1 ts2 ts3 ts4 : TypeSet .
11) vars t1 t2 t3 : Type .
12) vars sn1 sn2 sn3 sn4 : ServiceN .
13) vars srv1 srv2 srv3 : Service .
14) vars srvs1 srvs2 srvs3 : ServiceSet .
15) var e : Edge .
16) var es : Edges .
17) var nds : Nodes .
18) eq NamSer ( < sn1 , ts1 , ts2 > ) = sn1 .
19) eq NamSer ( ( sn1 : ts1 -> ts2 ) ) = sn1 .
20) ceq ( sn2 : ts3 -> ts4 ) ; srvs1 & | <
    sn1 , ts1 , ts2 > nds , es | =  ( sn2 :
    ts3 -> ts4 ) ; srvs1 & addE( | < sn1 ,
    ts1 , ts2 > nds , es | , link( < sn1 ,
    ts1 , ts2 > . < sn2 , Union( ts1 , ts3 )
    , Union(ts2 , ts4) > ) ) if ( Intersect(
    ts2 , ts3 ) =/= none )  /\ ( Inclus( ts4
    , ts2) == false ) /\ (not ( link( < sn1 ,
    ts1 , ts2 > . < sn2 , Union( ts1 , ts3 )
    , Union(ts2 , ts4) > ) in es) ) .
endm
```

Fig. 5. The functional Module «COMP-MODEL».

## 3.2 Construction of the composition plan

The second phase of the graph-based composition approach formalized in this work is to calculate the composition plans (i.e. services invocation sequence) using the graph model result of the first step and a user query. The inputs of this step are then: the services composition graph model and a request of the form *R.InputTypeList* → *R.requiredType-List*.

The module system «COMPOSITION-PLAN» presented in the figure 6 allows us to find from a services composition the compositions schemas completing a request. In this module, to add a node of the graph to the composition schema we must check two constraints: the node must have a maximum affinity (equal to 1) with the request and a link must exist with the last node in the schema.

Through the various modules presented in this section, we find that we have given a modular specification of the composition approach. So we can easily enrich this specification, and add other operations, sorts, equations, or even modules to specify different syntactic aspects which are not considered in our specification.

By the implementation of these modules in the Maude language, we obtain running specifications.

```
mod COMPOSITION-PLAN is
1)  including COMP-MODEL .
2)  protecting RAT .
3)  sorts Request COMP-PLAN .
4)  op _->_ : TypeSet TypeSet -> Request
    [ctor prec 25 ] .
5)  op<<_`,_`,_`,_>>:Cmp-Mod Request
    Nodes SetServiceN->COMP- PLAN .
6)  op aff : Request Node -> Nat [ctor] .
7)  op Card : TypeSet -> Nat [ctor] .
8)  vars ts1 ts2 ts1' ts2' ts3 ts4 ts5
    ts6 : TypeSet .
9)  vars sn1 sn2 : ServiceN .
10) vars srvs1 : ServiceSet .
11) vars invs : SetServiceN .
12) var e : Edge .
13) var es : Edges .
14) var nds nds' : Nodes.
15) var t1 : Type .
16) eq Card ( t1 ts1 ) = 1 + Card ( ts1 ) .
17) eq Card ( none ) = 0 [owise] .
18) eq aff (ts1 -> ts2, <sn1, ts3, ts4>)
    = Card( Intersect (ts2, ts4)) /Card
    (ts2).
19) crl [COMP-PL] :<< ( sn1 : ts1 -> ts2 ) ;
    srvs1 & | < sn1 , ts3 , ts4 > nds , es
    | , ts5 -> ts6 , niln , none >> => <<
    srvs1 & | < sn1 , ts3 , ts4 > nds ,es |,
    ts5 -> Remove( ts6 , ts2 ) , < sn1 , ts3
    , ts4 > , sn1 >> if Inclus(ts5 , ts3)
    /\ aff(ts5 -> ts6 ,< sn1,ts3 , ts4 >)==
    1 .
20) crl [COMP-PL] : << ( sn1 : ts1 -> ts2 ) ;
    srvs1 & | < sn1 , ts3 , ts4 > nds ,
    link( < sn1 ,ts3 , ts4 > . < sn2 , ts1'
    , ts2' > ) es | , ts5 -> ts6 , < sn2 ,
    ts1' , ts2' > nds' , sn2 invs >> => <<
    srvs1 & | < sn1 , ts3 , ts4 > nds ,
    link( < sn1 , ts3 , ts4 > . < sn2 , ts1'
    ,ts2' >) es | , ts5 -> Remove( ts6 , ts2
    ) , < sn1 , ts3 , ts4 > < sn2 , ts1' ,
    ts2' > nds' , sn1 ( sn2 invs ) >> if
    Inclus( ts5 , ts3 ) /\ aff ( ts5 -> ts6
    ,< sn1 , ts3 , ts4 >) == 1 .
endm
```

Fig. 6. The System Module «COMPOSITION-PLAN».

## 4. Case study

To better show the proposed formalization we present in this section a case study of the services collection «WEATHER-WS». Initially the set S of all services is composed of six atomic services S = {S1, S2, S3, S4, S5, S6}. Where each service has the inputs and outputs shown in Table 1.

Table 1: Inputs and outputs of the WEATHER-WS collection services.

| Atomic Service | InputTypeList | OutputTypeList |
|---|---|---|
| S1 | city | longitude, latitude |
| S2 | longitude, latitude | weather |
| S3 | zipecode | logitude, latitude |
| S4 | zipecode | weather |
| S5 | longitude, latitude, road | zipcode |
| S6 | city | zipcode |

```
mod WEATHER-WS is
1) ops city longitude latitude weather
   zipcode road : -> Type [ctor] .
2) ops s1 s2 s3 s4 s5 s6 : -> ServiceN[ctor].
3) op weather-ws :  -> ServiceSet [ctor] .
4) eq weather-ws = ( s1 : city ->
   longitude latitude ) ; ( s2 :
   longitude lati  tude -> weather );  (
   s3 : zipcode -> longitude latitude );
   ( s4 : zipcode -> weather ) ; ( s5 :
   longitude latitude road -> zipcode ) ;
   ( s6 : city -> zipcode ) .
endm
```

Fig. 7. The System module «WEATHER-WS».

The figure 7 shows the transformation of the WEATHER-WS collection services presented in Table 1 to a rewriting logic. More precisely, this transformation is done by declaring a system module that imports the generic module «COMPOSITION-PLAN» using the clause "extending", and the statement of manufacturing operations to identify in this case, the name of the collection (weather-ws), the names of atomic services (*S1, ..., S6*) and the types of exchanged data (city, longitude, ... road ). Finally, the last algebraic equation of the «WEATHER-WS» module groups all these elements.

To generate the graph-based model associated with the collection of WEATHER-WS services we must use the "*reduce*" Maude command while specifying the set S of all services and the initial node. Figure 8 shows an example of the running of this command. Figure 9 is a graph representation of the obtained results.

To show how to generate the various composition plans of a query, we introduce the example shown in the figure 10. This query has as input an element of type «city» and as required elements the set of types: «longitude», «latitude» and «weather». We must use the "*search*" Maude command while specifying the entire composition plan. The same figure shows the result obtained after the running.

## 5. Conclusions

In this work, we showed how the formalism of graphs can be used to improve the composition of web services and make it automatic. More precisely, we have proposed rewriting logic and its Maude language as a support for a graph-based approach for automatic composition of web services. The proposed model has made possible the exploration of different composition schemas as well as the formal analysis of service compositions. Our contribution has broadly followed two main steps:

- The first step consists of defining the graph model (implementation of the equational theory). This model offered the adequate semantic mechanisms specifying the constraints of connection between web services.
- The second step consists to generating the different schemas of possible compositions (composition plans) to accomplish a particular query. This phase is based on a mapping between the required elements and the provided elements of each node in the graph. This part constitutes the dynamic aspect of the formalization.

As an extension of this work, we aim to use the strategy technique of the Maude system to optimize the selection of the chosen services in the second phase of our formalization.

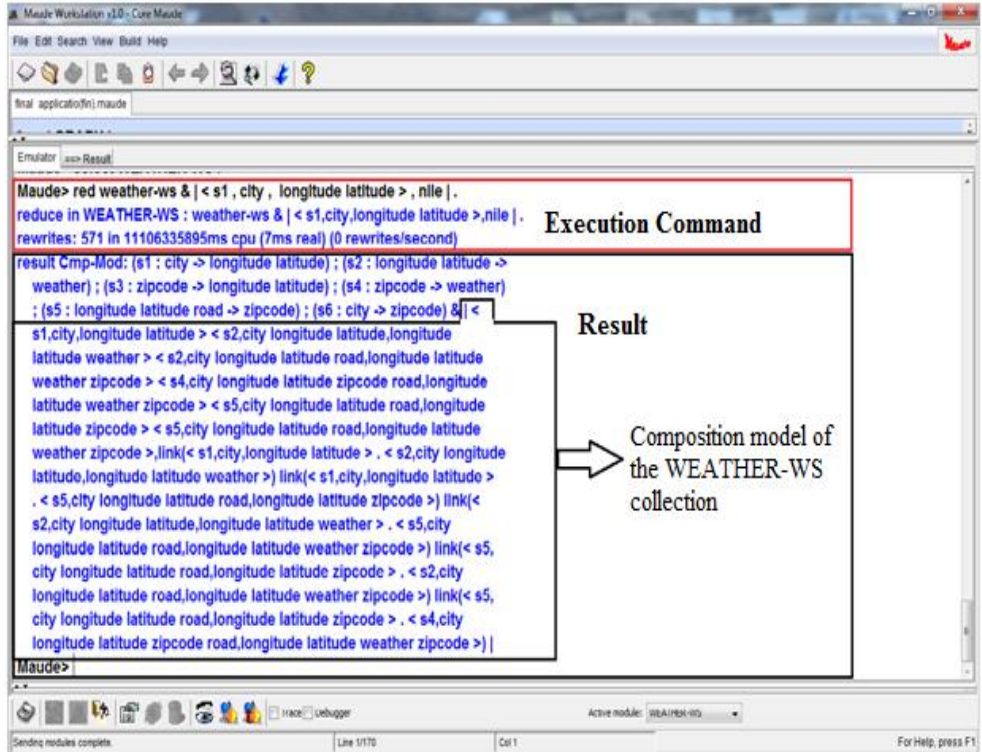

Fig. 8. Composition model of the WEATHER-WS collection.

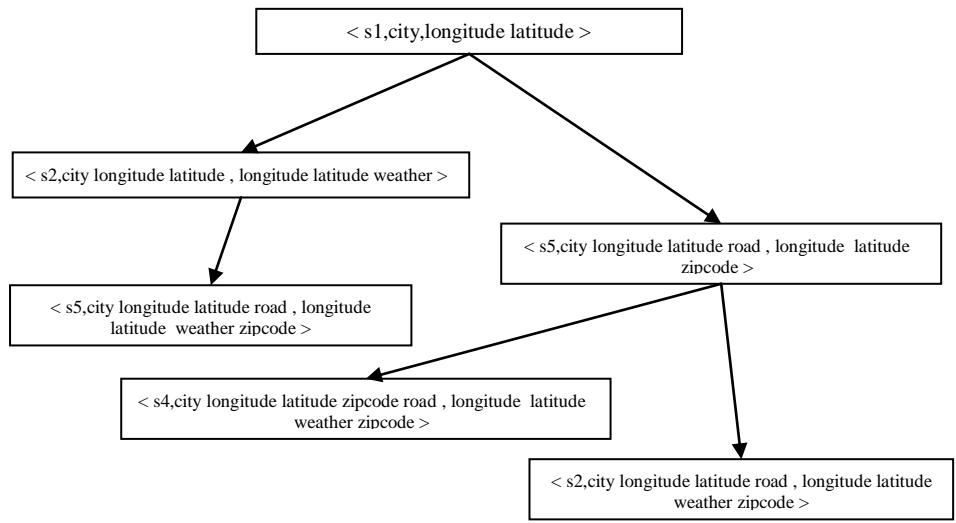

Fig. 9. Graph representation of the composition model.

```
                        \|||||||||||||||/
                      --- Welcome to Maude ---
                        /|||||||||||||||\
                    Maude 2.4 built: Dec 9 2008 20:35:33
                      Copyright 1997-2008 SRI International
                           Fri march 14 21:33:34 2014
    search in WEATHER-WS : << weather-ws & | < s1,city,longitude latitude >,nile |,
            c ity -> longitude latitude weather,niln,none >> =>*
            << C:Cmp-Mod,city -> none,nd:Nodes,S:SetServiceN >> .
```

**Solution 1 (state 5)**

states: 6  rewrites: 1389 in 4964211348ms cpu (10ms real) (0 rewrites/second)
C:Cmp-Mod --> (s3 : zipcode -> longitude latitude) ; (s4 : zipcode -> weather)
  ; (s5 : longitude latitude road -> zipcode) ; (s6 : city -> zipcode) & | <
  s1,city,longitude latitude > < s2,city longitude latitude,longitude
  latitude weather > < s2,city longitude latitude road,longitude latitude
  weather zipcode > < s4,city longitude latitude zipcode road,longitude
  latitude weather zipcode > < s5,city longitude latitude road,longitude
  latitude zipcode > < s5,city longitude latitude road,longitude latitude
  weather zipcode >,link(< s1,city,longitude latitude > . < s2,city longitude
  latitude,longitude latitude weather >) link(< s1,city,longitude latitude >
  . < s5,city longitude latitude road,longitude latitude zipcode >) link(<
  s2,city longitude latitude,longitude latitude weather > . < s5,city
  longitude latitude road,longitude latitude weather zipcode >) link(< s5,
  city longitude latitude road,longitude latitude zipcode > . < s2,city
  longitude latitude road,longitude latitude weather zipcode >) link(< s5,
  city longitude latitude road,longitude latitude zipcode > . < s4,city
  longitude latitude zipcode road,longitude latitude weather zipcode >) |
nd:Nodes --> < s1,city,longitude latitude > < s2,city longitude latitude,
  longitude latitude weather >  ⎬ *The set of nodes invoked in Cmp-Mod*
S:SetServiceN --> s1 s2  ⎬ *The set of services used in composition*

---

**Solution 2 (state 9)**
```
states: 10     rewrites: 1797 in 4964211348ms cpu (21ms real) (0
rewrites/second)
C:Cmp-Mod -->  ...............  .
nd:Nodes --> < s1,city,longitude latitude > < s2,city longitude latitude
road,longitude  latitude  weather  zipcode  >  < s5,city  longitude latitude
road,longitude latitude zipcode >
S:SetServiceN --> s1 (s5 s2)
```
**Solution 3 (state 10)**
```
states: 11     rewrites: 1905 in 4964211348ms cpu (204ms real) (0
rewrites/second)
C:Cmp-Mod -->..................
nd:Nodes --> < s1,city,longitude latitude > < s2,city longitude latitude,
longitude  latitude  weather  >  < s5,city longitude latitude road,longitude
latitude weather zipcode >
S:SetServiceN --> s1 (s2 s5)
```
```
No more solutions.
states: 12     rewrites: 1993 in 4964211348ms cpu (674ms real) (0
rewrites/second)
```

Figure 10. Composition model plan of the WEATHER-WS collection with running the "search" Maude command.

**Walid Berrouk** received his master in computer science from the Mentouri University in 2012. His research interests include web service composition and formal specification and verification.

**Ouanes AISSAOUI** is a Ph.D. student of Badji Mokhtar University in department of computer science. He received his master in computer science from the Badji Mokhtar University in 2011. His research interests include self-adaptive systems, software architecture and formal specification.